\documentclass[12pt,preprint]{aastex}
\usepackage{color}

\newcommand{\msun}{{\rm\,M_\odot}}
\newcommand{\Msun}{{\msun}}

\newcommand{\ct}{\citealt}

\begin{document}

\title{Orbital and Mass Ratio Evolution of Protobinaries Driven by Magnetic Braking}

\author{Bo Zhao\altaffilmark{1}, Zhi-Yun Li\altaffilmark{1}}
\altaffiltext{1}{University of Virginia, Astronomy Department, Charlottesville, USA}

\shortauthors{{\sc Zhao \& Li}}
\shorttitle{{\sc Protobinary Evolution Driven by Magnetic Braking}}

\begin{abstract}
The majority of stars reside in multiple systems, especially
binaries. The formation and early evolution of binaries is a
longstanding problem in star formation that is not yet fully 
understood. In particular, how the magnetic field observed in 
star-forming cores shapes the binary characteristics remains 
relatively unexplored. We demonstrate numerically, using an 
MHD version of the ENZO AMR hydro code, that a magnetic field 
of the observed strength can drastically change two of the 
basic quantities that characterize a binary system: the 
orbital separation and mass ratio of the two components. Our 
calculations focus on the protostellar mass accretion phase, 
after a pair of stellar ``seeds'' have already formed. We 
find that, in dense cores magnetized to a realistic level, 
the angular momentum of the material accreted by the 
protobinary is greatly reduced by magnetic braking. Accretion 
of strongly braked material shrinks the protobinary separation 
by a large factor compared to the non-magnetic case. The magnetic 
braking also changes the evolution of the mass ratio of 
unequal-mass protobinaries by producing material of low specific 
angular momentum that accretes preferentially onto the more 
massive primary star rather than the secondary. This is in 
contrast with the preferential mass accretion onto the secondary 
previously found numerically for protobinaries accreting from an unmagnetized 
envelope, which tends to drive the mass ratio towards unity. 
In addition, the magnetic field greatly modifies the morphology 
and dynamics of the protobinary accretion flow. It suppresses 
the traditional circumstellar and circumbinary disks that feed 
the protobinary in the non-magnetic case; the binary is fed 
instead by a fast collapsing pseudodisk whose rotation is strongly braked.   
The magnetic braking-driven inward migration of binaries from 
their birth locations may be constrained by high-resolution 
observations of the orbital distribution of deeply embedded 
protobinaries, especially with ALMA and JVLA.  

\end{abstract}

\section{Introduction}  
\label{intro}

%
% why study binaries? 
%
The majority of the Galactic field stars reside in multiple systems (\ct{DuquennoyMayor1991};
\ct{FischerMarcy1992}; \ct{Mason+1998}; \ct{ShatskyTokovinin2002};
\ct{GoodwinKroupa2005}; \ct{Raghavan+2010}; \ct{Janson+2012}); most of
those systems ($\sim 75\%$) are binaries (\ct{DuquennoyMayor1991}; \ct{TokovininSmekhov2002}). The multiplicity of young stellar objects is even higher (\ct{ReipurthZinnecker1993}; \ct{Mathieu+2000}; \ct{Duchene+2004}; \ct{Duchene+2007}), indicating that the formation of multiple systems, especially binaries, is a major, perhaps the dominant, mode of star formation. 

% 
% why study binary formation NOW? Orbital migration -- catchy phase!!
% 
How binaries form and evolve remains uncertain. This is particularly
true for the earliest, Class 0, phase, when the stellar seeds are
still deeply embedded inside a massive envelope. Although millimeter
interferometric observations such as \citet{Looney+2000} have
uncovered some Class 0 binaries, their number is still small, and the
distribution of orbital separation of such protobinaries is not well
constrained. \citet{Maury+2010} surveyed 5 Class 0 sources with PdBI
at sub-arcsecond resolution, and found a surprising result: only one
of the sources has a potential protostellar companion, and the
companion is $\sim 1,900$~AU away from the primary. Combining their
sample with that of \citet{Looney+2000}, \citet{Maury+2010} found that
none of the 14 Class 0 sources in the combined sample has a companion
with separation between $\sim 150-550$~AU, which is inconsistent with
the binary fraction of $\sim 18\%$ for Class I sources in the same
separation range (\ct{Connelley+2008}). A similar difference was found
in the CARMA survey of 6 Class 0 and 3 Class I sources in the Serpens
molecular cloud (\ct{Enoch+2011}). None of the Class 0 objects has any
detectable protostellar companion closer than ~2,000~AU (down to the
resolution limit $\sim 250$~AU), whereas one Class I object has a
companion at a projected distance of $\sim 870$~AU. Although the
statistical significance of the difference is still relatively low, it
brings into sharp focus the possibility that the distribution of
orbital separation of protobinaries may differ significantly from that
of their more mature counterparts. In other words, binaries may
migrate substantially from their birth locations, a situation somewhat
analogous to that inferred for hot Jupiters, although it is unclear
whether protobinaries would migrate inward or outward to fill the
apparent gap between $\sim 150-550$~AU. The direction (inward vs
outward) of protobinary orbital migration should be better constrained
observationally with large surveys of deeply embedded sources using
ALMA and JVLA. % large survey of Class 0 and I sources!! 

This paper focuses on one possible mechanism for moving the protobinaries away from their birth locations: magnetic braking. It may seem counter-intuitive that magnetic fields can change the binary orbit, because the magnetic forces do not act on the stars directly. However, the orbits of protobinaries are determined mainly by the angular momentum of the material to be accreted, which can be strongly affected, perhaps even controlled, by the magnetic field, through magnetic braking. The main goal of this paper is to quantify the extent to which the protobinary orbit is modified by a magnetic field of the observed strength.

%
% Magnetic field observations
% 
The strength of magnetic fields in star formation is usually measured by the dimensionless mass-to-flux ratio $\lambda$. It is the mass of a region divided by the magnetic flux threading the region in units of the critical value $(2\pi G^{1/2})^{-1}$ (\ct{NakamuraNakano1978}). For a sample of dense cores in the nearby dark clouds, \citet{TrolandCrutcher2008} inferred a mean value for $\lambda_{los}\sim 4.8$, based on the field strength and column density along the line-of-sight. Geometric corrections for projection effects should reduce this value by a factor of 2-3, yielding an intrinsic value of $\lambda$ of a few typically. Such a magnetic field is generally not strong enough to prevent the dense core from gravitational collapse and star formation. It is, however, strong enough to affect the angular momentum evolution of the collapsing core in general and binary formation in particular.

%
% Previous studies... main conclusions and limitation 
% motivate the focus of present investigation, protostellar accretion phase 
%

There have been several studies of the magnetic effects on binary formation (e.g., \ct{PriceBate2007}; \ct{HennebelleTeyssier2008}; \ct{Machida+2010}). These studies focused primarily on the classical mode of binary formation inside an isolated rotating core, through core fragmentation induced by an $m=2$ density perturbation (e.g., \ct{BossBodenheimer1979}, see \ct{Kratter2011} for a recent review). A general conclusion is that the fragmentation can be suppressed by a rather weak magnetic field if the core is only weakly perturbed. For example, \ct{HennebelleTeyssier2008} found that, for a density perturbation of $10\%$, the fragmentation is suppressed by a magnetic field corresponding to $\lambda=20$ (see their Fig.~3), much weaker than the observationally inferred field. In dense cores that are magnetized to a more realistic level (with $\lambda \sim$ a few), fragmentation can still occur, although a large amplitude perturbation is needed (\ct{PriceBate2007}; \ct{HennebelleTeyssier2008}). A limitation of the existing MHD calculations is that they are confined mostly to the pre-stellar phase of core collapse and fragmentation leading up to the formation of two binary seeds. How the binary seeds evolve during the subsequent protostellar mass accretion phase in the presence of a dynamically important magnetic field remains little explored. It is the focus of our investigation. 

% 
% outline and a taste of key results...How to bring up mass ratio naturally?
%
Our investigation of the magnetic effects on protobinary evolution will be carried out using an MHD version of the ENZO AMR hydro code that includes a sink particle treatment (\ct{BryanNorman1997}; \ct{O'Shea+2004}; \ct{WangAbel2009}; \ct{Wang+2010}). In \S~\ref{setup}, we discuss the setup for the initial binary seeds and the rotating magnetized protobinary envelope to be accreted by the seeds, as well as the numerical code used. In \S~\ref{result}, we present numerical results for the evolution of initially equal-mass binary seeds. We find that a magnetic field of the observed strength can shrink the protostellar orbit by more than an order of magnitude compared to the non-magnetic case. It also greatly changes the dynamics and morphology of the protostellar accretion flow near the binary. For initially unequal mass binary seeds, the issue of mass ratio evolution becomes important. 
In \S~\ref{massratio}, we follow the the evolution of unequal-mass protobinaries accreting from envelopes magnetized to different levels. We find that the well-known tendency towards equal mass in the non-magnetic case due to preferential mass accretion onto the secondary is weakened or even suppressed by magnetic braking. We summarize the main results and put them in context in \S~\ref{discussion}.

\section{Problem Setup}
\label{setup}

% motivation - define the scope of the problem, code, and core and sink treatment
As mentioned in the introduction, the focus of our investigation is on the protostellar accretion phase of binary formation, where the magnetic effects are least explored. We will therefore skip the pre-stellar phase of core evolution leading up to the production of a pair of binary seeds, and assume that, at the beginning of our calculations, the seeds are already formed and are ready to accrete from a rotating, magnetized envelope. The setup is similar in spirit to the influential work of \citet{BateBonnell1997} and \citet{Bate2000}, who studied the effects on binary properties of mass accretion from an unmagnetized envelope (see also \ct{Artymowicz1983}). We postpone a treatment of both the pre-stellar and protostellar phases of binary formation to a future investigation. In what follows, we describe the initial conditions for the envelope and binary seeds as well as the numerical code used for following the envelope collapse and protobinary accretion.  
%
%\subsection{Protobinary Envelope}
%\label{siscore}
%

Although protostellar envelopes are often observed to be irregular and filamentary (e.g., \ct{Tobin+2010}), it is instructive to model them simply, so that the basic effects can be illustrated as cleanly as possible. Since the formation and evolution of binaries involve complex dynamics that are challenging to simulate and interpret, it is useful to set up the calculations in such a way that the numerical results can be checked against analytic expectations. One way to achieve this is to start with a self-similar initial configuration for the envelope, with an $r^{-2}$ density distribution given by (\ct{Shu1977})
\begin{equation}\label{eq:sis}
\rho (r)= {{A c_s^2} \over {4 \pi G r^2}},
\end{equation}
where $c_s$ is the isothermal sound speed, and $A$ is an over-density parameter. The collapse of such an initial configuration is expected to remain self-similar (\ct{Shu1977}), and the self-similarity has proven to provide a powerful check on the numerically obtained solutions (see, e.g., Allen et al. 2003; \citeauthor{MellonLi2008} \citeyear{MellonLi2008, MellonLi2009}; Kratter et al. 2010; and discussion in \S~\ref{result}).  

We choose an over-density parameter $A=4$ (corresponding to a ratio of thermal to gravitational energy of $\alpha=3/(2A)=0.375$), so that the initial configuration is denser than the famous equilibrium singular isothermal sphere by a factor of 2. The mass enclosed within any radius $r$ can be integrated as,
\begin{equation}\label{eq:mass}
M(r)={{A c_s^2} \over G } r.
\end{equation}
For an adopted core radius $R=10^{17}$~cm and isothermal sound speed $c_s=0.2$~km/s (corresponding to a temperature of $\sim 10$~K), the above equation yields a total core mass $M_{tot}=1.2\Msun$ and an average free-fall time $t_{ff}\approx 88$~kyr.

We generalize the isothermal configuration to include both rotation
and magnetic fields. To preserve the self-similarity, the rotation
speed cannot depend on radius but can have an angular dependence
(\ct{Allen+2003}). A convenient choice is $v_{\phi}=v_0$~sin~$\theta$
(where $\theta$ is the polar angle measured from the rotation axis),
which ensures that the angular speed is finite on the axis. We pick
$v_0=c_s$, in order to have as large a rotation speed as possible, so
that the binary can be well resolved, especially for strongly
magnetized cases where the binary separation is reduced by a large factor compared to the non-magnetic case; even faster, supersonic, rotation may produce undesirable shocks. The adopted rotation profile corresponds to a ratio of rotational to gravitational energy $\beta=(v_0/c_s)^2/(3A) \approx 0.083$, somewhat higher than used in other works (e.g., \ct{Machida+2010}). Nonetheless, it is still within the range inferred by \citet{Goodman+1993} from NH$_3$ observations of dense cores.

We choose an initially uni-directional magnetic field along the rotation axis (or the $z$-axis in the simulations), with the field strength decreasing away from the axis as $1/\varpi$ (where $\varpi$ is the cylindrical radius), such that the mass-to-flux ratio is constant spatially. To avoid singularity at the origin, we soften the profile to $1/(\varpi + r_h)$, so that 
\begin{equation}
B_z(\varpi)={ {A c_s^2} \over {\sqrt{G} \lambda} } {1 \over {\varpi}+r_h},
\end{equation}
where $\lambda$ in the denominator is the dimensionless mass-to-flux ratio of the envelope in units of the critical value $(2{\pi}G^{1/2})^{-1}$, and the parameter $r_h$ is defined below. Even though dense cores typically have $\lambda$ of a few (see discussion in \S~\ref{intro}), we will consider a much wider range of $\lambda=$2, 4, 8, 16 and 32, as well as the non-magnetic case ($\lambda=\infty$), so as to capture any trend that may exist in the protobinary properties as the field strength increases gradually. 

%
%\subsection{Seed binary}
%\label{binaryseeds}
%

To study the protobinary evolution, we follow \citet{BateBonnell1997}
and \citet{Bate2000} and insert two equal-mass ``seeds'' near the center of
the protostellar envelope at the beginning of the calculation;
non-equal mass binaries will be treated separately in
\S~\ref{massratio}. We assume that each of the binary seeds has a
small initial mass of $0.05 \Msun$. To determine the initial binary
separation, we assume that the binary seeds get both their masses and
orbital angular momentum from a sphere of radius $r_h = 8.30 \times 
10^{15}$~cm in the initial envelope that contains $0.1\Msun$. In 
other words, we assign both the mass and angular momentum inside $r_h$ 
to the binary seeds. The initial orbital angular momentum is thus 
\begin{equation}\label{eq:am}
L_b = L(r_h) = \int\limits^{r_h}_0\int\limits^{\pi}_0\int\limits^{2\pi}_0 {\rho v_0 \sin\theta \varpi r^2 \sin\theta \,dr d\theta d\phi} = {{A c_s^2 v_0} \over {3G}} r_h^2,
\end{equation}
which, for circular orbits, yields an initial binary separation of 
\begin{equation}
a = {{16 A^2 c_s^4 v_0^2 r_h^4} \over {9 (G M_b)^3}}=3.687 \times 10^{15}\ \rm{cm} \approx 246\ \rm{AU}, 
\end{equation}
where $M_b=0.1~\Msun$ is the total mass of the binary.

To follow the envelope accretion and protobinary evolution, we use an MHD version (\ct{WangAbel2009}) of the ENZO adaptive mesh refinement hydro code (\ct{BryanNorman1997}; \ct{O'Shea+2004}). It incorporates a sink particle treatment (\ct{Wang+2010}). The magnetic field is evolved with a conservative MHD solver that includes the hyperbolic divergence cleaning of \citet{Dedner+2002}. The MHD version of the code is publicly available from the ENZO website at {\it http://code.google.com/p/enzo/}. It has been used to follow successfully the formation and evolution of magnetized galaxies (\ct{WangAbel2009}) and star clusters (\ct{Wang+2010}), as well as single star formation in magnetized dense cores (\ct{Zhao+2011}). 

We carry out simulations with periodic boundary conditions inside a box of length $5\times 10^{17}$~cm on each side, which is significantly larger than the protobinary envelope (of radius 
$10^{17}$~cm). The region outside the envelope is filled with a uniform isothermal medium of the same density (and temperature) as that at the outer boundary of the envelope ($\rho_a \approx 1.91 \times 10^{-19}$~g~cm$^{-3}$), so that the pressures are initially balanced across the boundary. To speed up computation, we adopt a relatively coarse base grid of $64^3$, but allow for 8 levels of refinement, with a smallest cell size of about 2.0~AU. 

As usual, we adopt a barotropic equation of state (EOS) that mimics the isothermal EOS at low densities and adiabatic EOS at high densities: 
\begin{equation}
P(\rho) = \rho c_s^2 \left[1 + \left({\rho\over\rho_{crit}}\right)^{2/3}\right],
\end{equation}
with a critical density $\rho_{crit}=10^{-13}$~g~cm$^{-3}$ for the transition between the two regimes. 

% Sink treatment and sink test!!

We treat the binary stars as sink particles. The sink particle treatment is described in detail in \ct{Wang+2010}. Briefly, each particle accretes according to a modified Bondi-Hoyle formula (see \ct{Ruffert1994}). New sink particles are created at the center of those over-dense cells that violate the Jeans criterion at the highest refinement level, i.e. when $\rho_{cell} > \rho_{J} = {\pi \over G} ({c_s \over {N \Delta x}})^2 = 7.90 \times 10^{-14} g \cdot cm^{-3}$, where we have used a Truelove's (\ct{Truelove+1997}) safety number $N=16$ and a finest cell size $\Delta x=3.05 \times 10^{13}$~cm. Newly created sink particles are subject to merging, which is controlled by two parameters: the merging mass $M_{merg}=0.01\Msun$ and merging distance $l_{merg}=5 \times 10^{14}$~cm (for details, see \ct{Wang+2010}). These values are chosen to eliminate artificial particles and to maximize computation efficiency. Our main results are insensitive to these parameters as long as they are reasonably small.

We have tested the sink particle accretion algorithm against known semi-analytic solutions of the collapse of (non-magnetic) singular isothermal sphere (\ct{Shu1977}) and found good agreement. For example, the collapse of our chosen initial configuration (Eq.~[\ref{eq:sis}] with $A=4$) yields, after some initial adjustment, a constant mass accretion rate of $1.06\times 10^{-5}$~$\Msun$~yr$^{-1}$, which matches Shu's result almost exactly. 

In a magnetized medium, we let the sink particle accrete only the mass but not the magnetic field; the field is left behind. This treatment is a crude representation of the decoupling of the magnetic field from the matter expected at high densities (\ct{Nakano+2002}; see discussion in \ct{Zhao+2011}). 

\section{Magnetic Braking and Protobinary Orbital Evolution}
\label{result}

\begin{figure}
\epsscale{1.1}
\plottwo{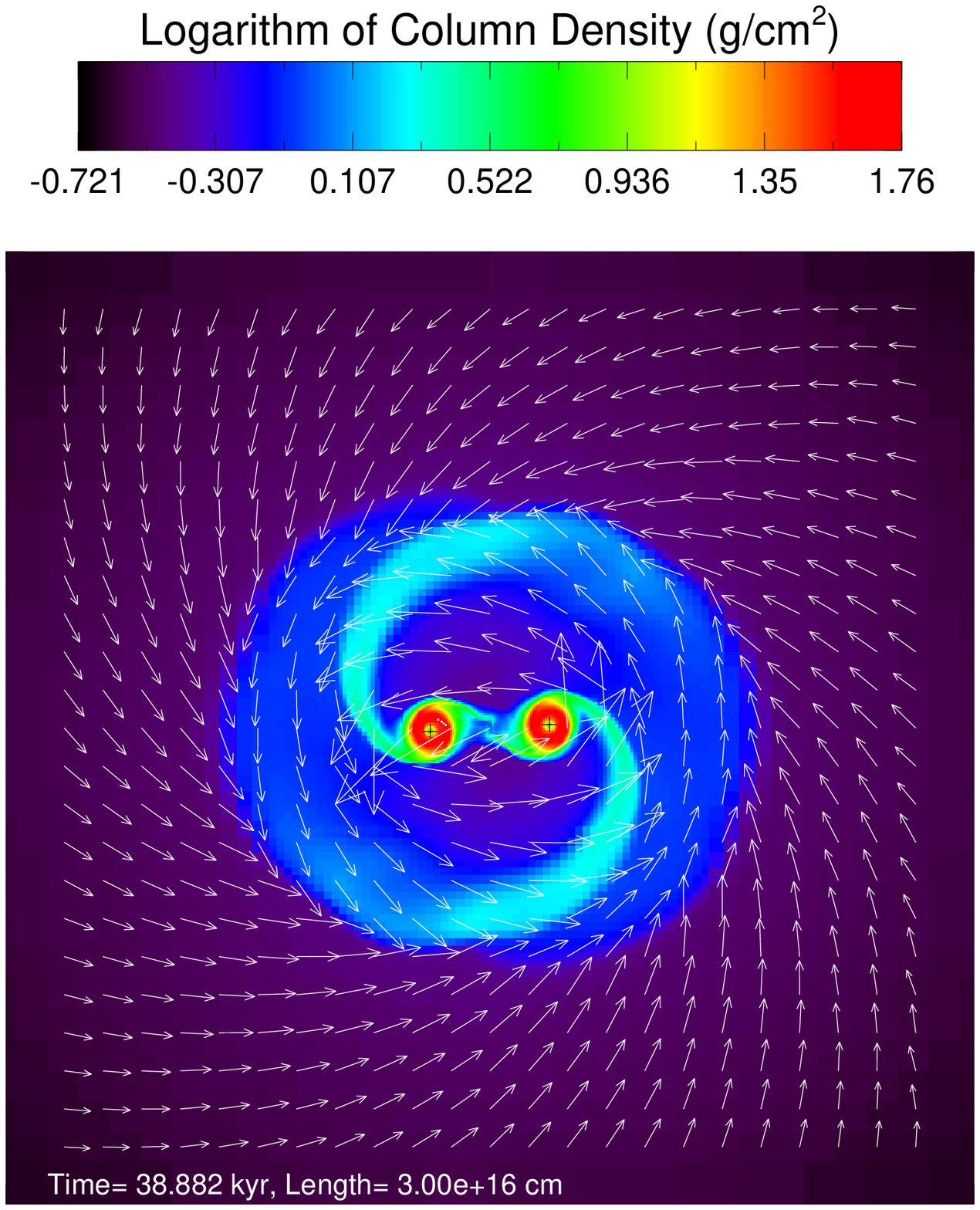}{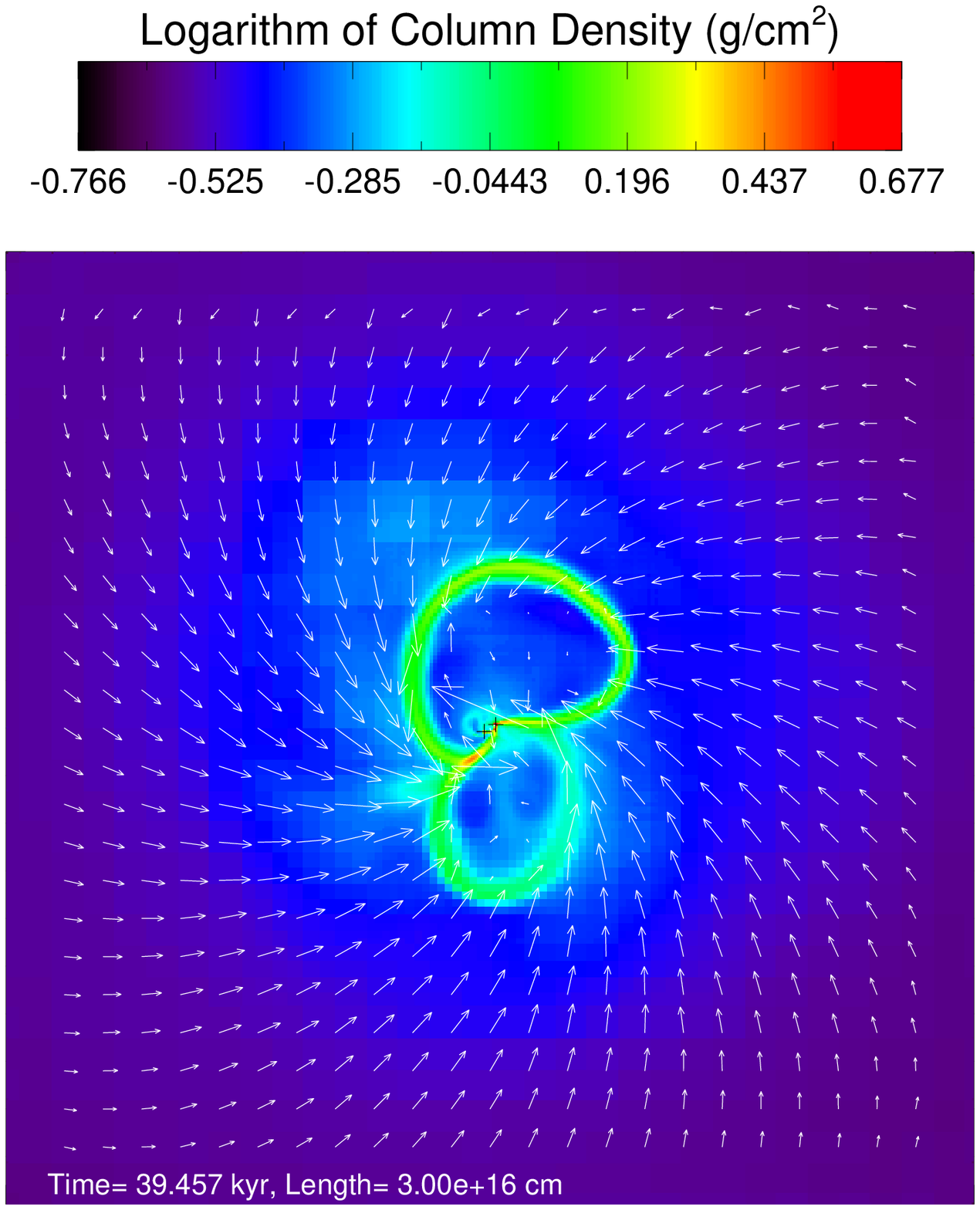}
\caption{Distribution of the logarithm of the column density $\Sigma$ (in $g\cdot cm^{-2}$ ) along z-direction and velocity field on the equatorial plane. Both the HD (left panel) and $\lambda=2$ (right panel) case are at $t\approx 39$~kyr. The well-defined circumstellar and circumbinary disks in the former are replaced by two magnetically dominated lobes in the latter. The binary stars are marked by crosses.}
\label{figcomp}
\end{figure}

Before discussing detailed quantitative results, we first describe the
qualitative effects of the magnetic field on the (initially
equal-mass) protobinary orbit by contrasting the two extreme cases:
the case without a magnetic field (the HD or hydro case hereafter), 
and the strongest field
case ($\lambda=2$, which is consistent with available Zeeman
measurements, see \ct{TrolandCrutcher2008}). Both simulations reached a rather late time 
($t\gtrsim 80$~kyr, comparable to the free-fall time at the average density), when the binary seeds have finished many orbits around each other. In Fig.~\ref{figcomp}, we show the snapshots of the two cases near the middle point of the simulation, at $t\approx 39$~kyr. They look strikingly different. In the HD case, there are two well separated stars, each surrounded by a circumstellar disk. The disks are surrounded by a well-defined circumbinary structure, with two prominent spiral arms embedded in it. The circumstellar disks and circumbinary structure are similar to those found in previous non-magnetic calculations, such as \citet{Bate2000} and \citet{Hanawa+2010}. They are the result of the conservation of angular momentum in the rapid collapsing region of the envelope, and redistribution of angular momentum close to the binary through gravitational torque. 

These well-known features are completely absent in the strongly
magnetized $\lambda=2$ case. They are replaced by two irregular lobes,
which were studied in detail in \citet{Zhao+2011}; they are the
so-called ``magnetic decoupling-enabled structures'' (or DEMS for
short) produced by the magnetic flux decoupled from the matter that
enters the stars (sink particles). The DEMS are magnetically
dominated, low-density, expanding regions. They surround the
protobinary, whose separation is much smaller than that of the HD case
(see Fig.~\ref{figcomp}). Clearly, the magnetic field has greatly modified not only the circumstellar and circumbinary structures, but also the binary orbit. In the following subsections, we will discuss these modifications more quantitatively. 

\subsection{Binary Separation and Angular Momentum-Mass Relation}
\label{separation}

\subsubsection{Hydro Case: Checking against Expectations}
% make a case that HD is consistent with expectation quantitatively!

\begin{figure}
\epsscale{1.0}
\plotone{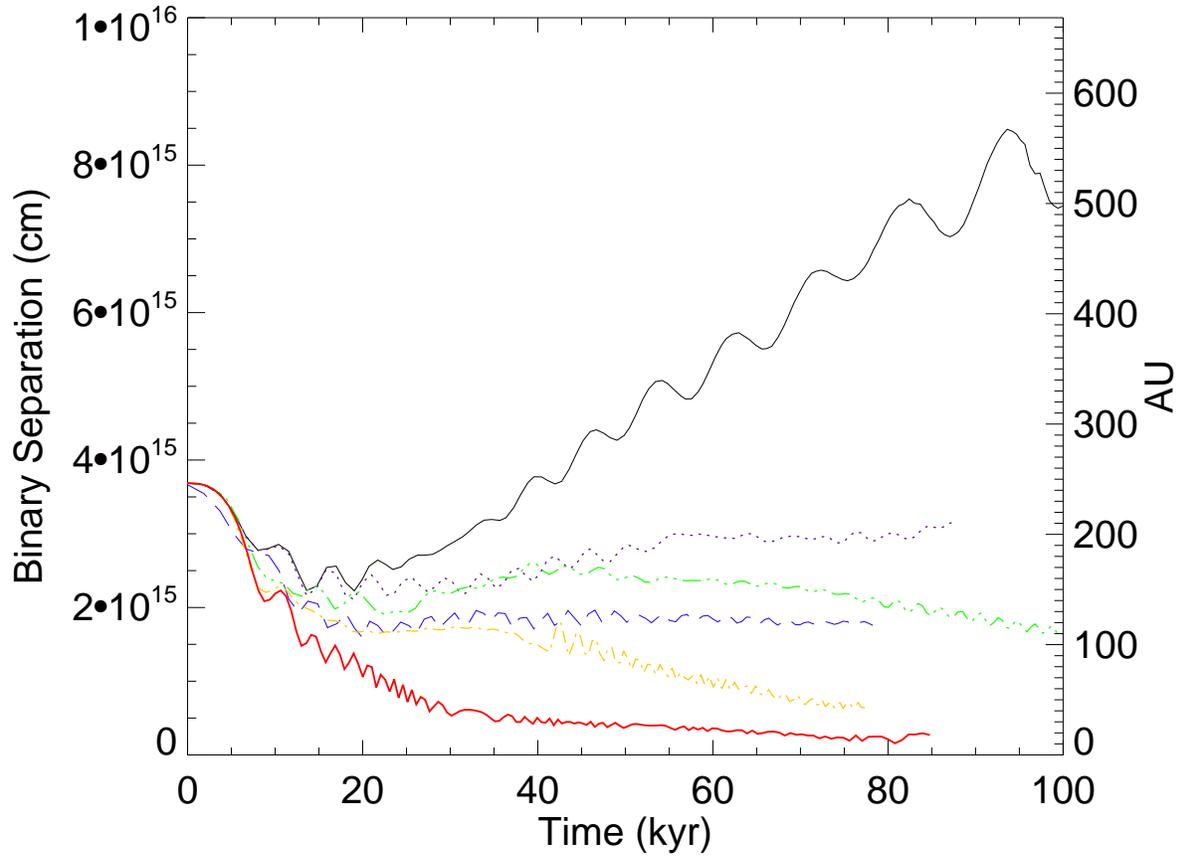}
\caption{Evolution of binary separation with time for HD (black solid), $\lambda=32$ (purple dotted), $\lambda=16$ (blue long-dashed), $\lambda=8$ (green dash-dot-dot-dotted), $\lambda=4$ (yellow dash-dotted), and $\lambda=2$ (red thick solid) cases. Note the large difference between the HD and the realistically magnetized 
($\lambda=2$ and perhaps 4) cases. 
}
\label{figbinsep}
\end{figure}

We begin our quantitative discussion with binary separation. It is
shown 
as a function of time in
Fig.~\ref{figbinsep} for all initially equal-mass simulations. 
The evolution of the binary
separation is particularly interesting in the HD case. It decreases 
initially for about 20~kyr, before increasing almost linearly with 
time\footnote{The wiggles on the curve is due to small orbital 
eccentricity (of order $10\%$ or less) excited during the protobinary 
evolution. We will postpone a detailed study of the magnetic effect 
on eccentricity to a future investigation.}. The linear increase 
is a natural consequence of the initial 
protostellar envelope configuration chosen, which is self-similar 
over a range of 
radii (excluding the regions close to the center and the outer edge). 
The configuration is expected to collapse self-similarly, with the 
binary separation increasing linearly with time (for the same
reason that the size of the expansion wave in the well-known inside 
out collapse of a singular isothermal sphere increases linear with 
time; \ct{Shu1977}), after some initial adjustment. The decrease in binary separation during the initial 
adjustment comes about because we have put all of the angular 
momentum of the material inside a small sphere into the orbit of the initial binary seeds. This leads to an overestimate of the initial binary orbital angular momentum (and thus the separation) because a fraction of the angular momentum should be left behind in the circumstellar disks and circumbinary structure. 
We have experimented with binary seeds formed out of a smaller sphere 
that have smaller initial masses and separation. They exhibits a similar 
linear growth of separation with time after a shorter adjustment period. 
In other words, the protobinary reaches the expected self-similar state 
more quickly. The agreement of the numerical result with expectations 
gives us confidence that the ENZO-based AMR code can treat the 
protobinary evolution problem properly. 

\begin{figure}
\epsscale{1.0}
\plotone{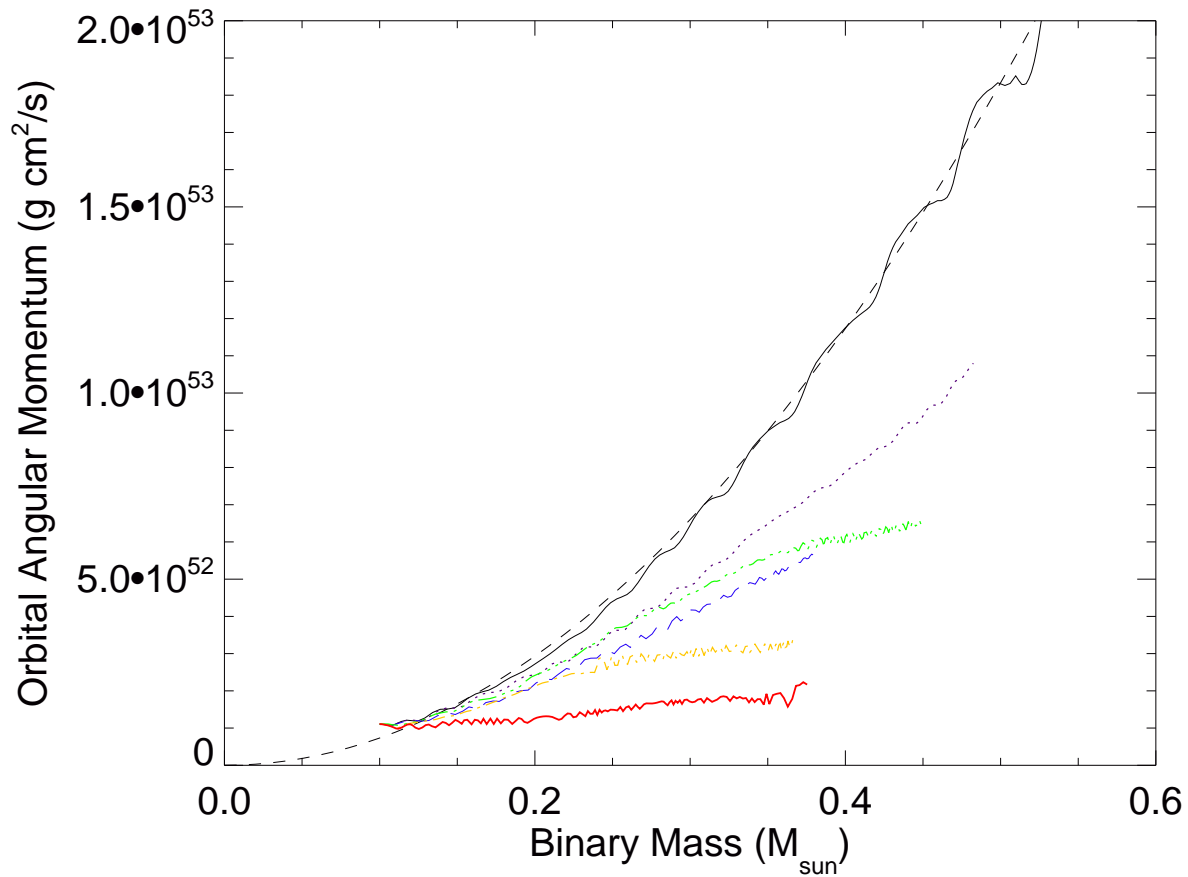}
\caption{The relation between the orbital angular momentum and total
  mass of the protobinary. The dash line is the the analytical
  prediction from Eq.~\ref{eq:LvsMrel} (reduced by a correction factor
  of $\sim 1.5$). The different curves are: HD (black solid), $\lambda=32$
  (purple dotted), $\lambda=16$ (blue long-dashed), $\lambda=8$ (green dash-dot-dot-dotted), $\lambda=4$ (yellow dash-dotted), and $\lambda=2$ (red thick solid).}
\label{figbinLM}
\end{figure}

Another check on the numerical solution comes from the expected
scaling between the orbital angular momentum and total mass of 
the protobinary. The scaling can be obtained as follows. 
We assume that the gas in our singular isothermal envelope accretes 
shell by shell onto the binary. The total mass $M(r)$ and total 
angular momentum $L(r)$ for a sphere of radius $r$ are given, respectively,  
by Eq.~(\ref{eq:mass}) and 
\begin{equation}\label{eq:angmom}
L(r) = {A c_s^2 v_0\over 3 G} r^2.
\end{equation} 
If all of the mass and angular momentum of the material within the sphere 
are accreted onto the binary, then the orbital angular momentum $L_b$ 
and binary mass $M_b$ must be related through 
\begin{equation}\label{eq:LvsMrel}
L_b(M_b)={{2 G v_0} \over {3 A c_s^2}} M_b.
\end{equation}
This relation is shown as the dashed line (scaled down by a factor 
of $\sim 1.5$) in Fig.~\ref{figbinLM}, together with the angular 
momentum-mass (or $L_b$-$M_b$) relations obtained in all of our simulations. 
Again, the HD case is particularly noteworthy. Its $L_b-M_b$ relation closely 
matches the analytical prediction from eq.~(\ref{eq:LvsMrel}), except
for a correction factor of $\sim 1.5$. The correction is to be expected 
because the mass and orbital angular momentum of the protobinary
at any given time does not come from a region of perfectly spherical
shape (assumed in deriving the above equation). More slowly rotating 
material near the rotation axis can collapse more quickly onto the 
binary than that near the equator, lowering the actual angular 
momentum of the binary relative to its mass. In any case, the broad 
agreement between the numerical result on the $L_b$-$M_b$ relation and the
analytical expectation lends further credence to the correctness of
the hydro simulation.

\subsubsection{Magnetic Effect on Protobinary Orbit}

Fig.~\ref{figbinsep} shows a general trend that the binary separation 
decreases with increasing magnetic field strength. The difference 
is especially striking at late times, when the separation increases 
with time for the HD case but stays roughly constant or decreases 
for the magnetized cases. By the time $t\approx 80$~kyr, the
protobinary separation in the HD case reaches $\sim 500$~AU, which 
is much larger than that in the $\lambda=4$ ($\sim 50$~AU) and 
$\lambda=2$ ($\sim 10$~AU) case. There is
little doubt that a realistic magnetic field (corresponding to a
dimensionless mass-to-flux ratio of a few, see Troland \& Crutcher 
2008 and discussion in \S~\ref{intro}) can shrink the protobinary 
orbit by a large factor. 

The reduction in binary separation is related to a decrease in the 
orbital angular momentum of the system. This is shown explicitly in 
Fig.~\ref{figbinLM}. 
There is a general trend for the orbital angular
momentum to decrease with increasing magnetic field strength. When 
the mass of the stellar seeds quadruples from 0.1 to
$\sim$0.4~$\Msun$, the orbit angular momentum 
in the HD case increases by more than an order of magnitude, whereas
that in the strongly magnetized $\lambda=2$ case increases by merely 
a factor of 2. In the latter case, there is a large amount of mass 
added to the protobinary but relatively little angular momentum. Since 
the binary 
separation $a$ is sensitive to the orbital angular momentum $L_b$ 
(i.e., $a\propto L_b^2$ for a fixed binary mass $M_b$), even a 
relatively modest change in the angular momentum would lead to a 
significant change in the separation. 
 
We conclude from Figs.~\ref{figbinLM} and \ref{figbinsep} that the
main magnetic effect on the protobinary evolution is to reduce its 
orbital angular momentum compared to the non-magnetic case, which 
in turn leads to a tighter orbit. This result may appear puzzling 
at the first sight, because the magnetic forces do not act directly 
on the binary seeds. However, they do act on the material to be 
accreted by the seeds. By changing the angular momentum of
such material, the magnetic field can greatly affect, 
perhaps even control, the orbital evolution of the protobinary. In 
the next subsection, we explore in some detail the mechanism through 
which the magnetic field shrinks the protobinary orbit. 

\subsection{Magnetic Braking and Angular Momentum Removal from 
Protobinary Accretion Flow}
\label{angularmomentum}

It is well-known that magnetic fields interact strongly with fluid 
rotation, through magnetic braking. If the infalling material has a 
large fraction of its angular momentum removed prior to its arrival 
at the binary seeds, it would add mass but relatively little angular 
momentum to the protobinary system. As a result, the binary 
separation would increase less rapidly with time compared to the 
hydro case; it may even decrease with time if the magnetic braking 
is strong enough.  

\begin{figure}
\epsscale{0.9}
\plotone{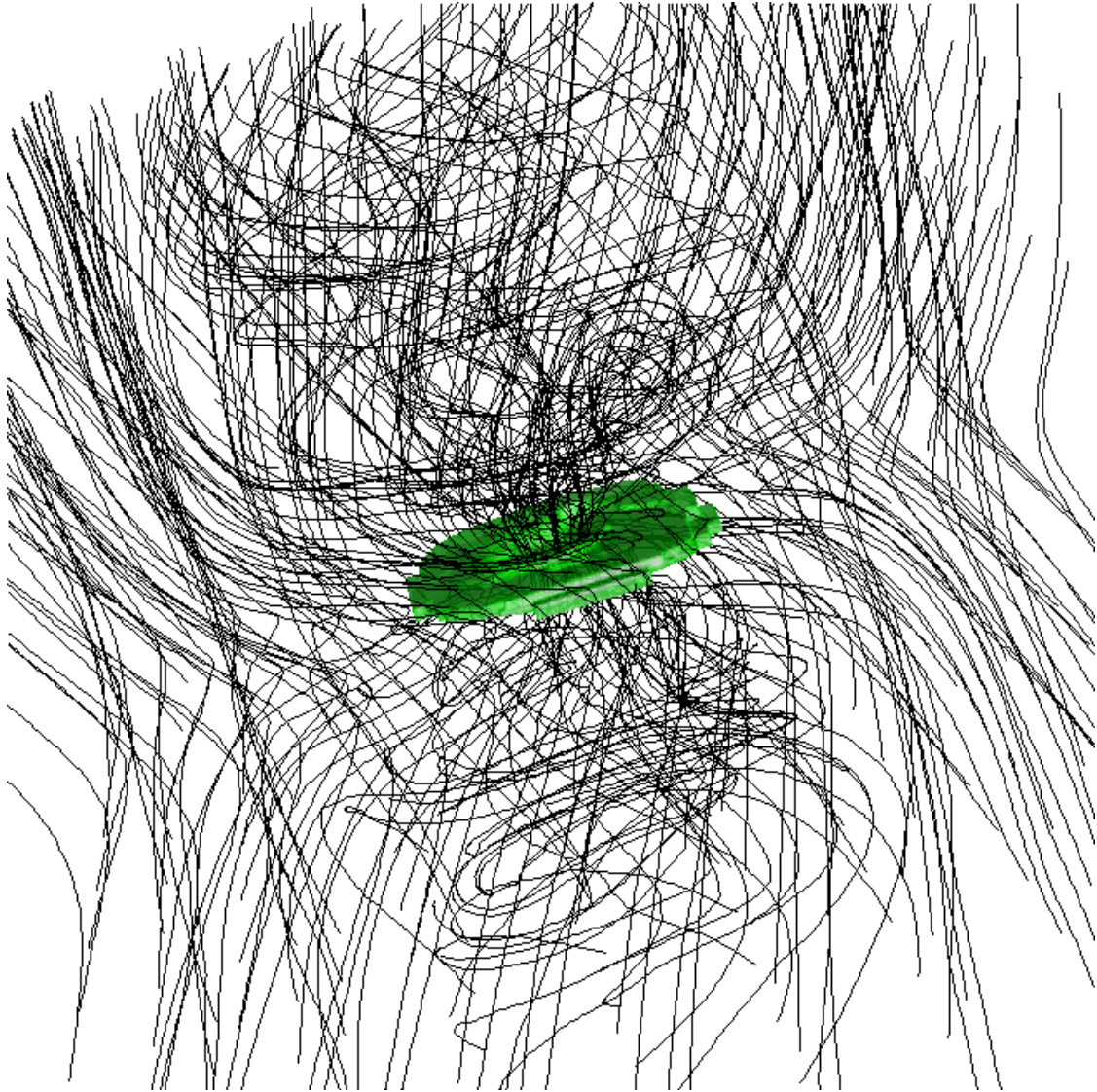}
\caption{3D view of the magnetic field lines and an iso-density
  surface in the inner part of the protobinary accretion flow at the
  representative time $t\approx 39$~kyr, showing the field twisting
  that is the smoking gun of magnetic braking. The plotted region has
  a dimension of $5\times 10^{15}$~cm. }
\label{braking}
\end{figure}

The presence of magnetic braking can be seen directly from Fig.~\ref{braking},
where we plot the field lines at the representative time $t\approx 
39$~kyr for the $\lambda=2$ case. The initially straight field lines 
are twisted by the fluid rotation into helical shape, especially in the polar regions. The magnetic tension force associated with the twist 
acts back on the fluid, braking its rotation. 

To quantify the strength of the magnetic braking, let us consider a
finite volume $V$ with surface $S$. Inside this volume, the total
magnetic torque relative to the origin (from which a radius vector 
${\bf r}$ is defined) is 
\begin{equation}
{\bf N}_{m}={1 \over {4\pi}} \int[{\bf r} \times ((\nabla \times {\bf B}) \times {\bf B})]\,dV,
\end{equation}
where the integration is over the volume $V$. Typically, the magnetic
torque comes mainly from the magnetic tension rather than pressure
force. The dominant magnetic tension term can be simplified to a 
surface integral (\ct{MatsumotoTomisaka2004})
\begin{equation}
{\bf N}_t={1 \over {4\pi}} \int ({\bf r} \times {\bf B})({\bf B} \cdot d{\bf S}),
\end{equation}
over the surface $S$ of the volume. This volume-integrated magnetic
torque is to be compared with the rate of angular momentum advected
into the volume through fluid motion, 
\begin{equation}
{\bf N}_a=-\int \rho({\bf r} \times {\bf v})({\bf v} \cdot d{\bf S}),
\end{equation}
which will be referred to as the advective torque below. 

Since the initial angular momentum of the protobinary envelope is 
along the $z$-axis,
we will be mainly concerned with the $z-$component of the magnetic
and advective torque, 
\begin{equation}
N_{t,z} =  {1 \over {4\pi}} \int (x B_y - yB_x)({\bf B} \cdot d{\bf S})
\end{equation}
and 
\begin{equation}
N_{a,z} = -\int \rho (x v_y - y v_x)({\bf v} \cdot d{\bf S}),
\end{equation}
which can change the $z-$component of the angular momentum $L_z$
within the volume $V$. 

\begin{figure}
\epsscale{1.0}
\plotone{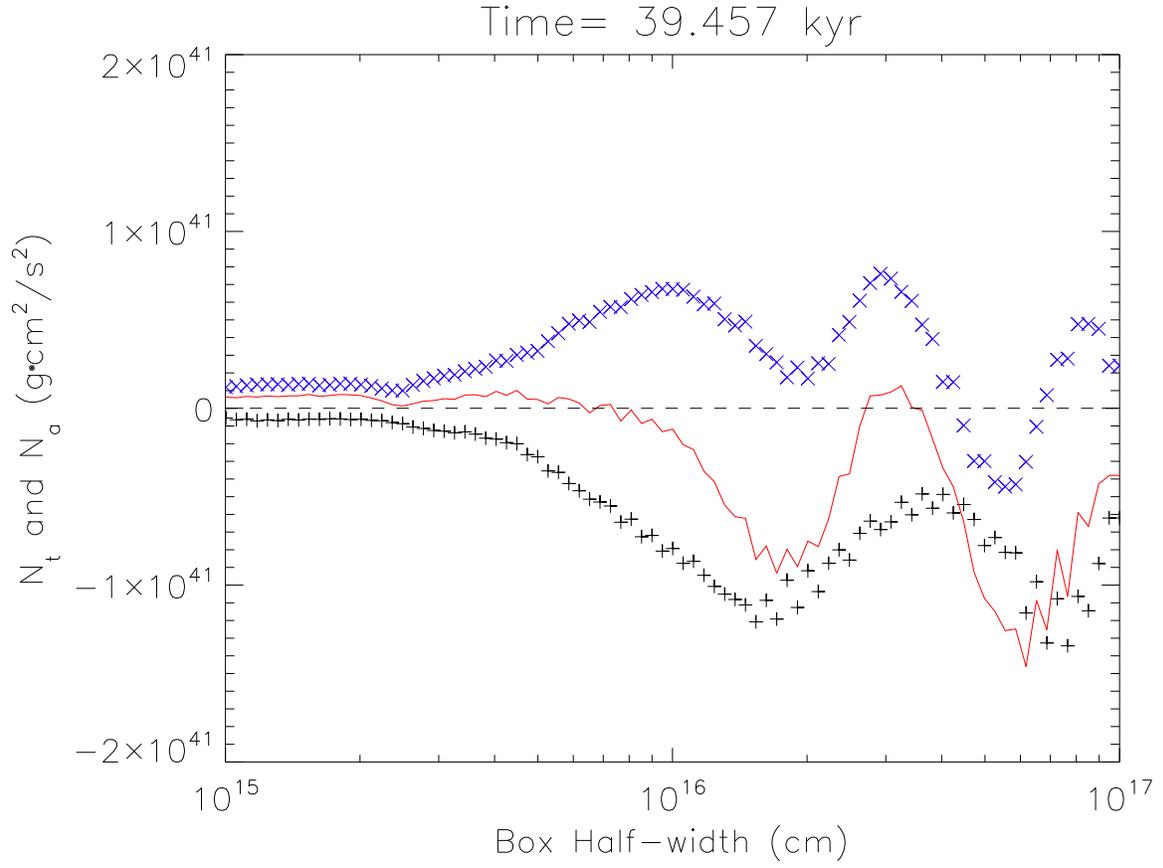}
\caption{The magnetic (black `+') and advective (blue `x') torque and the sum
  of the two (red) for cubic boxes of different half-width $b$ for 
the $\lambda=2$ case, at a representative time $t\approx 39$~kyr. 
A positive torque increases the angular momentum within a volume whereas
a negative one decreases it. }
\label{figtorq}
\end{figure}

As an example, we show in Fig.~\ref{figtorq} the distributions of the
magnetic and advective torques $N_{t,z}$ and $N_{a,z}$ for cubic boxes 
of different sizes that are 
centered at the origin, at the representative time $t\approx 39$~kyr 
for the $\lambda=2$ case. As expected, the volume-integrated magnetic 
torque is negative for boxes of most sizes; it removes angular
momentum from the material inside the boxes through magnetic braking. 
At the time shown, the braking torque is particularly large outside 
$\sim 10^{16}$~cm from the origin, reaching absolute values of order 
$10^{41}$~g~cm$^2$s$^{-2}$ or larger. To appreciate how large
this torque is, we note that the orbital angular momentum of the 
protobinary in the hydro case is $\sim 4.4\times 10^{52}$~g~cm$^2$s$^{-1}$ 
around the same time. This angular momentum would be removed by the above 
magnetic torque on a time scale of $13$~kyr, much shorter than the envelope  
collapse time. Given the magnitude of the magnetic torque, it is not 
surprising that the orbital angular momentum (and thus the separation) 
of the protobinary can be reduced significantly. 

The magnetic torque changes with time, however. At the time shown
in Fig.~\ref{figtorq}, the magnetic torque overwhelms the advective 
torque for boxes of most sizes, leading to a net decrease of the
angular momentum with time for the material in these boxes. This may not happen at other times. To evaluate the
accumulative effects of the magnetic and advective torque over time, 
we define for each box of half width $b$ 
\begin{equation}
L_{t,z}(b,t)=\int_0^t N_{t,z}(b,t^\prime) dt^\prime 
\end{equation}
and
\begin{equation}
L_{a,z}(b,t)=\int_0^t N_{a,z}(b,t^\prime) dt^\prime 
\end{equation}
which are the amount of the $z-$component of the angular momentum
inside the box changed by the magnetic and convective torque,
respectively, up to time $t$. These two quantities are to be
compared with the actual change of the angular momentum inside each 
box between time $t$ and $t=0$,
\begin{equation}
\Delta L_z(b,t) = L_z(b,t)-L_z(b,t=0).
\end{equation} 

\begin{figure}
\epsscale{0.8}
\plotone{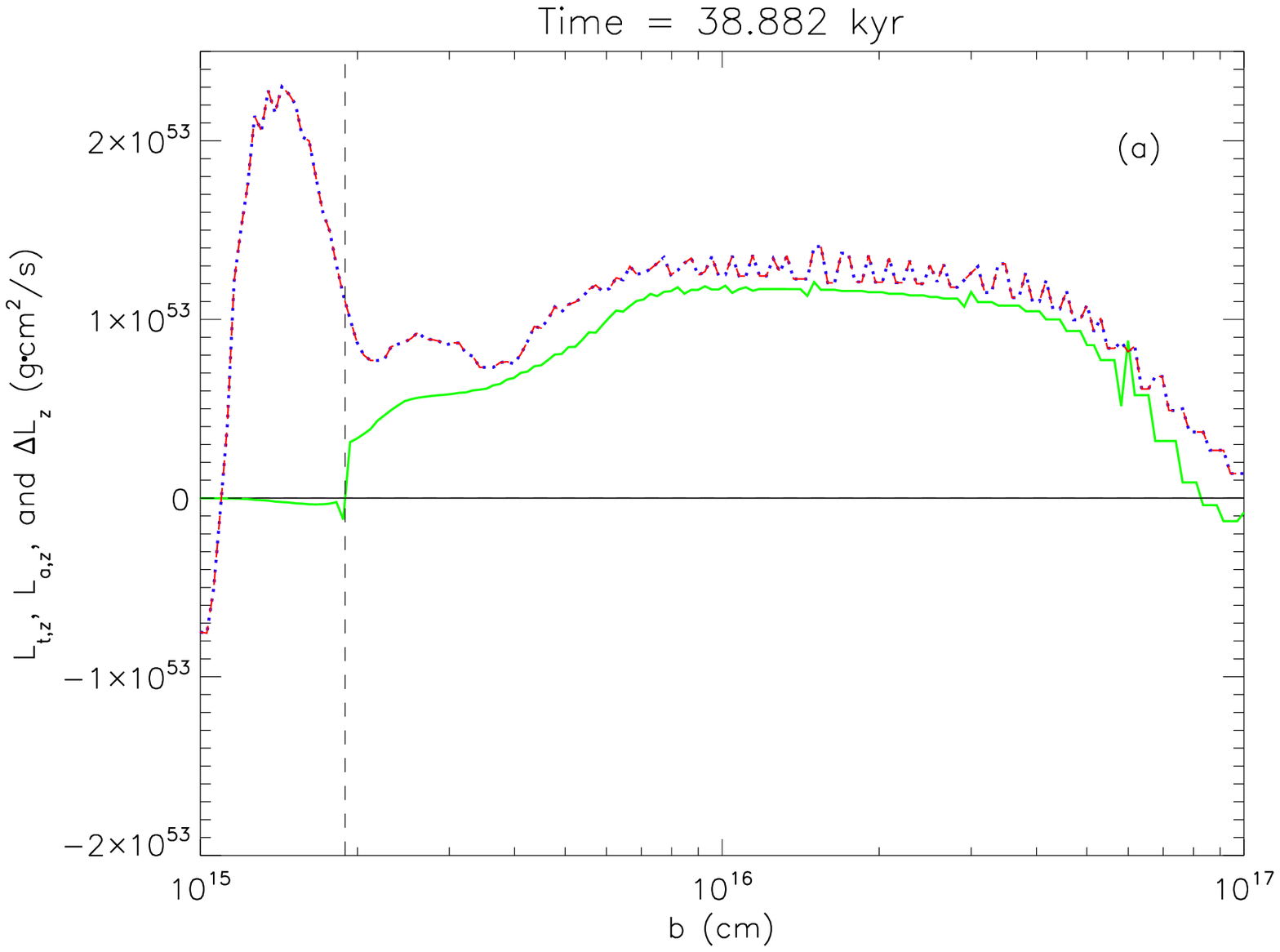}
\plotone{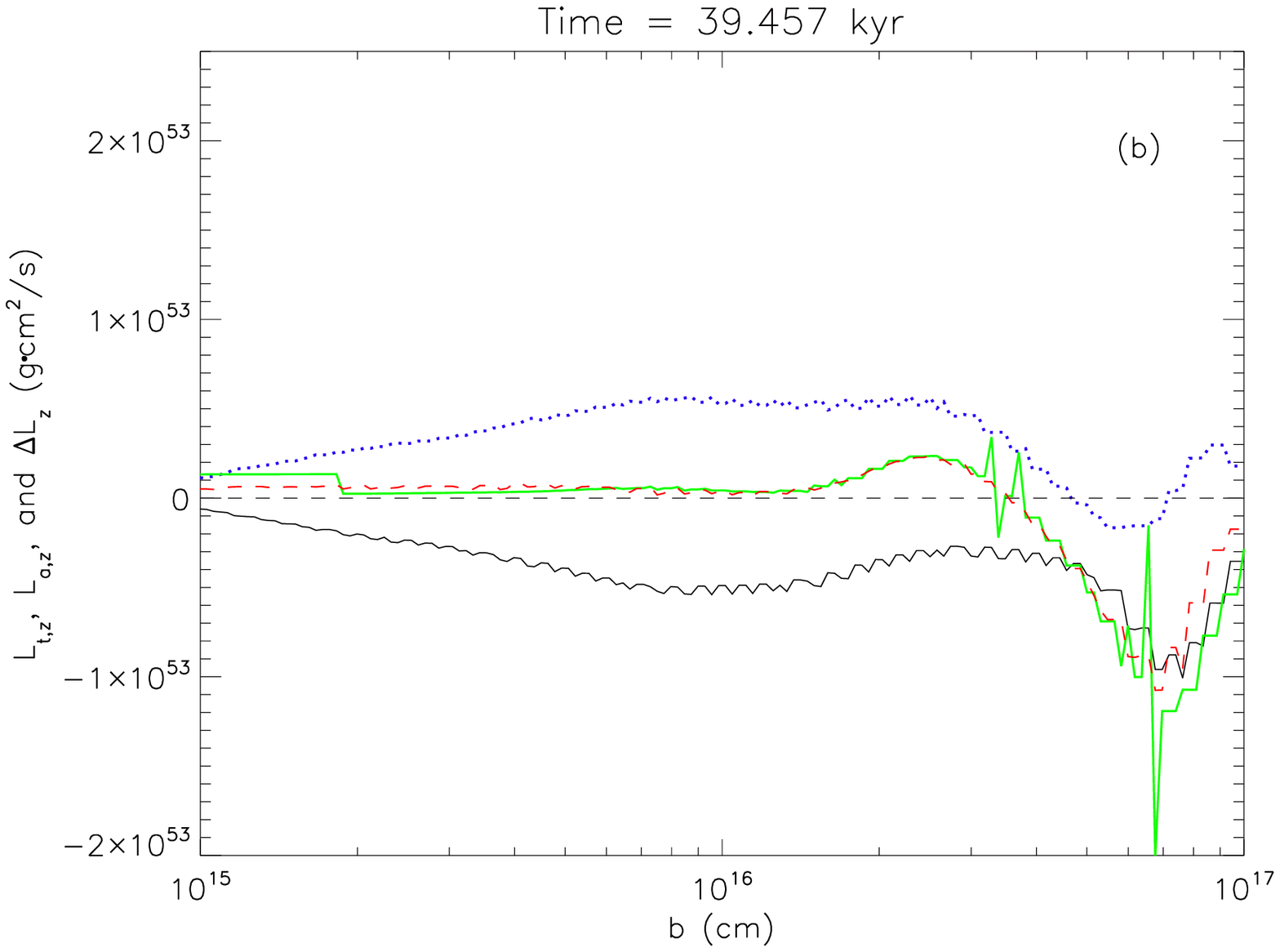}
\caption{The integrated magnetic and advective torques $L_{t,z}$ (black solid), $L_{a,z}$ (blue dotted), and their sum $L_{t,z} + L_{a,z}$ (red dashed) together with the actual angular momentum change $\Delta L$ (green thick solid) for cubic boxes of different half-width $b$. The HD case is shown in panel (a) with zero $L_{t,z}$ and the $\lambda=2$ case is shown in panel (b), both at a similar time $t\approx 39$~kyr. The vertical dashed line in panel (a) indicates the approximate position of the sink particles, whereas in panel (b) such particle position lies below $10^{15}$~cm.}
\label{figint}
\end{figure}

Barring a significant gravitational torque, one expects $\Delta
L_z(b,t)$ to be close to $L_{a,z}(b,t)$ in the hydro case, because
fluid advection should be the main channel for angular momentum 
change. This is indeed the case, as illustrated in panel (a) of Fig.
~\ref{figint}, where both $\Delta L_z(b,t)$ and $L_{a,z}(b,t)$ 
are plotted as a function of the box size $b$ at the representative 
time $t\approx 39$~kyr. The change in angular momentum 
$\Delta L_z(b,t)$ does follow closely the advected angular momentum 
for boxes of sizes larger than $\sim 10^{16}$~cm. For boxes of smaller 
sizes, there is significantly more angular momentum advected into 
a box than the actual angular momentum change in it, indicating that a 
good fraction of the angular momentum advected into the box is 
transported back out, presumably by the gravitational torques 
associated with the spiral arms that are visible in the left panel 
of Fig.~\ref{figcomp}. The gravitational torques may also be 
responsible for the small excess of $L_{a,z}(b,t)$ over $\Delta 
L_z(b,t)$ for the larger boxes. 

The situation is very different in the presence of a relatively 
strong magnetic field, as shown in panel (b) of 
Fig.~\ref{figint}. Plotted are $L_{t,z}(b,t)$, $L_{a,z}(b,t)$, 
and $L_{t,z}(b,t)+L_{a,z}(b,t)$ together with $\Delta L_z(b,t)$ 
as a function of box size at $t\approx 
39$~kyr for the $\lambda=2$ case. The total change of the 
angular momentum due to the magnetic and advective 
torque acting on the boundary of a box, $L_{t,z}(b,t)+L_{a,z}(b,t)$,  
is very close to the 
actual change in angular momentum inside the box, $\Delta L_z(b,t)$, 
indicating that any additional torques (such as gravitational torques), 
if present, play a relatively minor role in angular momentum 
transport. This is not surprising, because the prominent spiral 
arms of the hydro case are disrupted by the magnetic field 
completely. More importantly, the sum $L_{t,z}(b,t)+L_{a,z}(b,t)$ 
has a magnitude much smaller than $\vert L_{t,z}(b,t)\vert$ and 
$\vert L_{a,z}(b,t)\vert$ individually, which means that most of 
the angular momentum advected into a box is removed by the 
magnetic braking, leaving little net angular momentum change 
in it. This is unequivocal evidence that the magnetic 
braking in the $\lambda=2$ case controls the angular momentum 
evolution of the protostellar accretion flow, which in turn 
shapes the orbit of the protobinary.  

The effect of magnetic braking is even stronger than indicated by panel (b) of Fig.~\ref{figint}. This is because $L_{a,z}(b,t)$ is 
the net angular momentum advected into a box, i.e., the sum of the 
positive angular momentum advected into the box $L_{a,z}^+(b,t)$ 
and the negative angular momentum advected out of the box 
$L_{a,z}^-(b,t)$. In the $\lambda=2$ case, the magnetic braking 
drives an outflow, which produces a negative angular momentum 
$L_{a,z}^-(b,t)$ that is not much smaller in magnitude than 
$L_{a,z}^+(b,t)$. In other words, the angular momentum advected by 
infall into a box is larger than that shown in Fig.~\ref{figint}, 
and most of this larger angular momentum is removed by both 
the magnetic braking itself and the braking-induced outflow.  

\begin{figure}
\epsscale{1.1}
\plottwo{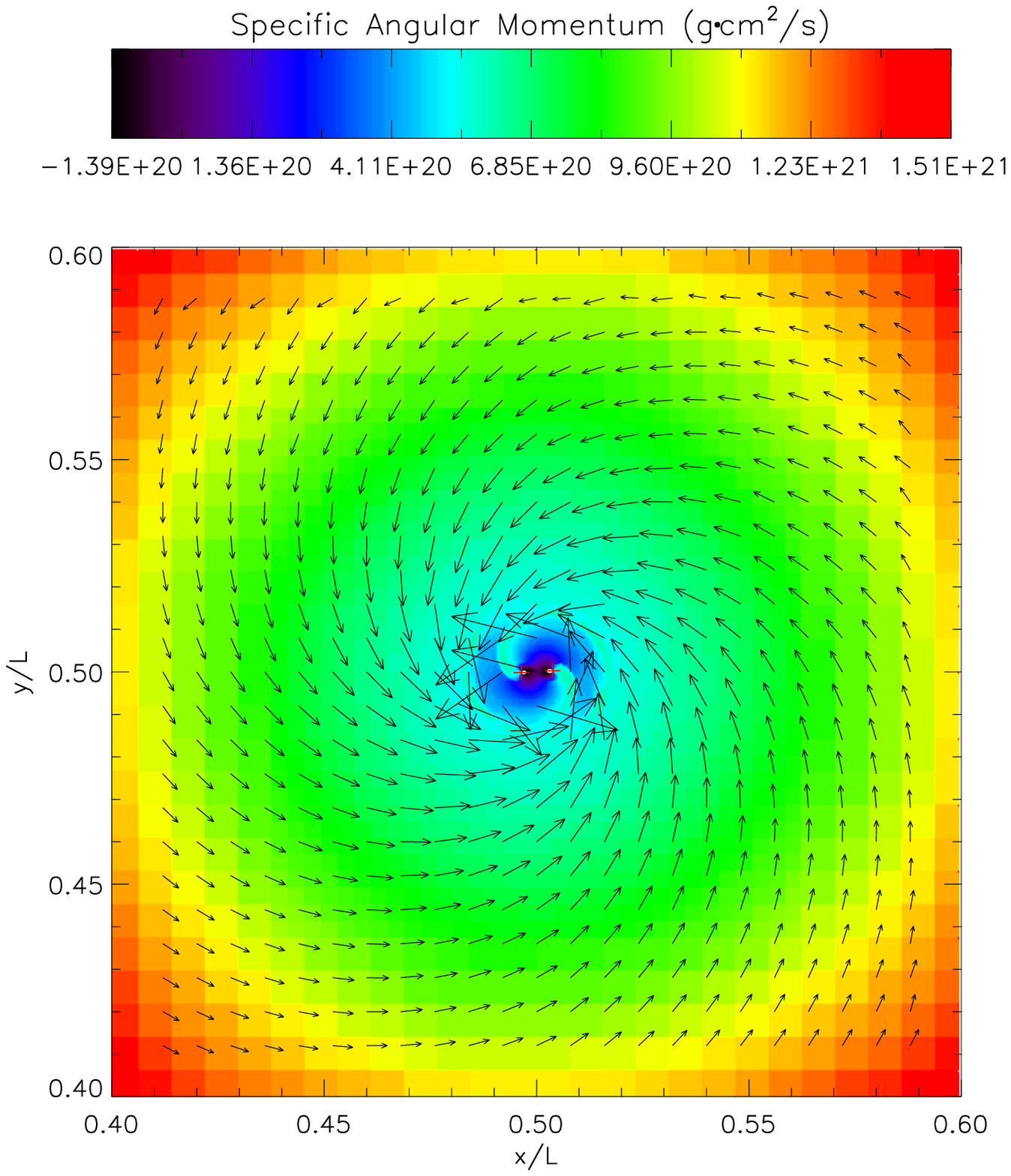}{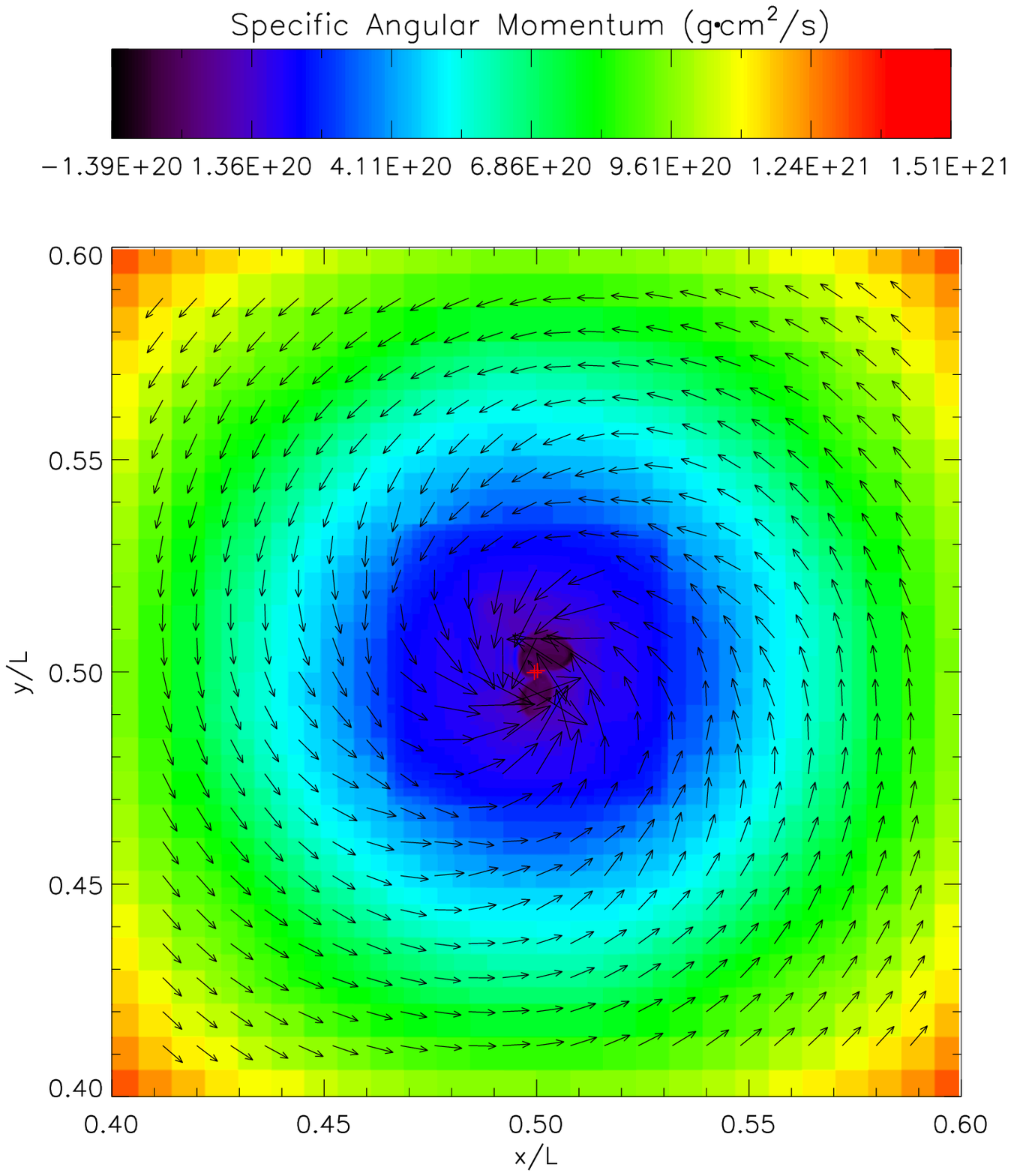}
\caption{Snapshot of the gas specific angular momentum (in $g\cdot cm^2\cdot s^{-1}$ with logarithm scale) on the equatorial plane of the inner accretion flow for the HD case (left panel) and the $\lambda=2$ case (right panel) at $t\approx 39$~kyr, showing the strong braking of the material to be accreted by the protobinary in the magnetized case compared to the HD case. The arrows are velocity vectors, and the crosses mark stars. Only the central region of $10^{17}$~cm on each side is plotted.}
\label{figbrake}
\end{figure}

To illustrate the effect of magnetic braking more visually, we 
plot in Fig.~\ref{figbrake} the distribution of specific angular
momentum on the equatorial plane for the hydro and $\lambda=2$ case 
at the representative time $t\approx 39$~kyr. It is clear that, 
for the hydro case, the specific angular momentum is roughly 
constant within a dimensionless radius of $\sim 0.05$ (or 
$\sim 2.5\times 10^{16}$~cm), indicating that the collapsing material 
has a more or less conserved angular momentum before it is 
accreted by the protobinary. The relatively large specific 
angular momentum of the accreted material is what drives the 
binary separation to increase. In the $\lambda=2$ case, the 
specific angular momentum of the material to be accreted is 
much smaller; it is reduced by twisted field lines (see Fig.~\ref{braking}) as the 
material falls toward the binary. It is the accretion of 
the severely braked, low angular momentum material that drives 
the protobinary closer with time.

\subsection{Circumstellar and Circumbinary Structures}
\label{disks}

We have already seen from Fig.~\ref{figcomp} that a realistic 
magnetic field of $\lambda=2$ changes the circumstellar and
circumbinary structures of the HD case greatly: the prominent features
of the HD case, two well-defined circumbinary disks and a circumbinary 
disk with two prominent spiral arms, are replaced by two low density 
lobes, which are filled with the magnetic flux decoupled from the matter that has been accreted onto the binary, i.e., the DEMS ({\ct{Zhao+2011}; see also Seifried et al. 2011; Joos et al. 2012; \ct{Krasnopolsky+2012}). These strongly magnetized structures present an obstacle to mass accretion onto the protobinary. Although the DEMS tend to be less prominent for weaker magnetic fields, they still dominate the face-on view of the $\lambda=4$ and 8 cases, especially at late times. An implication is that, for protostellar envelopes magnetized to a realistic level (with $\lambda\sim$ a few), the protobinary mass accretion does not follow the traditional path: from the envelope to a circumbinary disk to circumstellar disks to individual stars. Rather, the envelope material collapses directly close to the stars typically, and be accreted along azimuthal directions not occupied by the DEMS. 

Another magnetic effect on the circumbinary environment is illustrated in Fig.~\ref{fluffy}, where we display both the edge-on and face-on view of the $\lambda=2$ case at a relatively early time $t=18$~kyr. The edge-on view shows that the protobinary is surrounded by two large, mostly expanding, regions (one each above and below the equator) that are absent in the HD case. Their dynamics are dominated by the rotationally twisted magnetic fields (see Fig.~\ref{braking}). It is in these regions that most of the angular momentum extracted magnetically from the material that falls into the stars is stored. Such bipolar expanding regions are seen in many magnetized core collapse simulations (for an early example, see \ct{Tomisaka1998}). 
They block the protobinary accretion over most of the solid angle, and force the accretion to occur mainly through a flattened, equatorial structure --- a circumbinary pseudodisk (\ct{GalliShu1993}). 
\begin{figure}
\epsscale{0.9}
\plotone{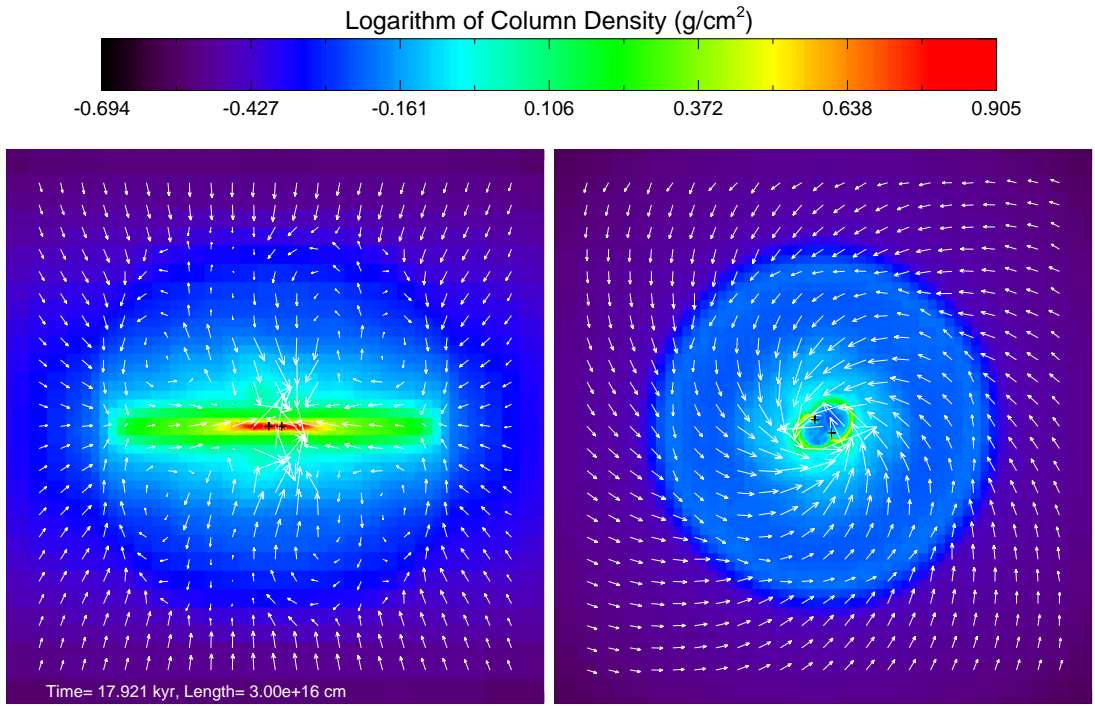}
\caption{Left panel: edge-on view of the column density in the
  $\lambda=2$ case at $t=18$~kyr, showing two slowly expanding polar  
regions sandwiching an infalling, circumbinary pseudodisk in the 
equatorial region. Right panel: face-on view of the same structure,
  showing that the circumbinary pseudodisk is not rotationally
  supported; it collapses rapidly. }
\label{fluffy}
\end{figure}

The pseudodisk is also clearly visible in the face-on view of the
system in the right panel of Fig.~\ref{fluffy}, as a nearly circular
region of enhanced column density. Unlike the circumbinary disk in the HD case (see the left panel of Fig.~\ref{figcomp}), the pseudodisk is not rotationally supported. It falls supersonically inward, and the infall can be seen in the velocity fields shown in Fig.~\ref{fluffy}. Another difference is that the circumbinary pseudodisk does not have prominent spiral arms. This is not surprising, because the pseudodisk is strongly magnetized, which makes gas compression more difficult. The rapid infall also leaves little time for the spirals to develop. The lack of spiral arms is consistent with our earlier result that the angular momentum transport in the $\lambda=2$ case is dominated by magnetic braking rather than gravitational torque. 

\subsection{Protobinary Mass Growth} 

Despite the differences in the morphology and dynamics of the
circumbinary environment with and without a magnetic field, the stars
grow at remarkably similar rates in all cases 
(of order $\sim 10^{-6} \Msun /$yr). This is shown in Fig.~\ref{figacc}, where the mass of each component of the protobinary is plotted as a function of time, for different degrees of magnetization. The reason for the similarity is that the mass accretion rate is determined mainly by the dynamics on the envelope scale: the material collapsing down from the envelope will find a way to the stars sooner or later, irrespective of the details of the circumbinary environment. In the HD or weakly magnetized cases, there is more material parked in the circumstellar and circumbinary disks, making the mass accreting rate onto the stars lower. However, this effect is smaller than that from the retardation of envelope collapse by magnetic tension in the stronger field cases. As a result, the protostellar mass accretion rate is somewhat lower in the stronger field cases, although not by a large factor. 
 
\begin{figure}
\epsscale{1.0}
\plotone{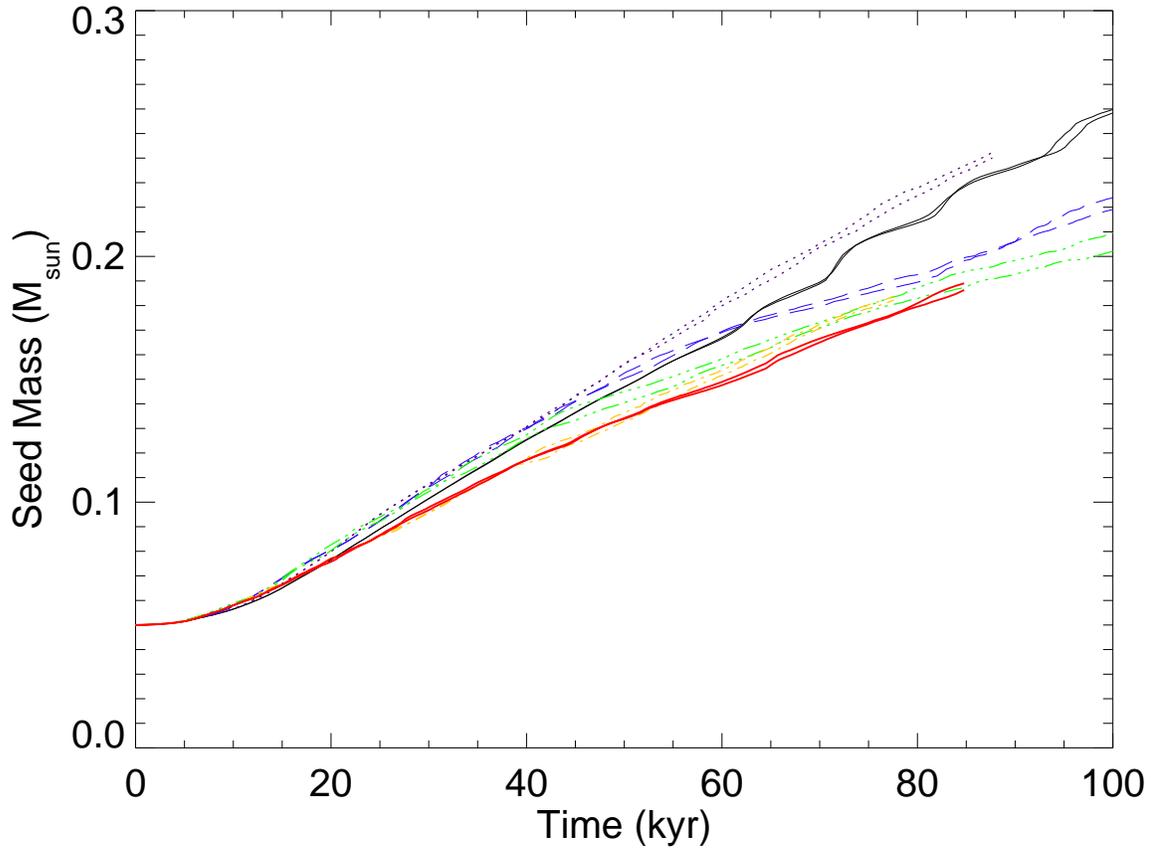}
\caption{Stellar mass (in solar units) growth of the initially
  equal-mass binary system for HD (black solid), $\lambda=32$ (purple dotted), $\lambda=16$ (blue long-dashed), $\lambda=8$ (green dash-dot-dot-dotted), $\lambda=4$ (yellow dash-dotted), and $\lambda=2$ (red thick solid) cases. Note that, in all cases, the two stars stay roughly equal-mass at all times.}
\label{figacc}
\end{figure}

It is reassuring that the masses of the two initially equal-mass
components stay nearly the same for all cases, despite the fact that
the inner protobinary accretion flow can become rather disordered at
times, especially for strongly magnetized cases. The magnetic field
does not appear to change much the rate of mass accretion by one component relative to the other, although the situation is drastically different for initially unequal mass protobinaries, as we demonstrate next. 

\section{Magnetic Braking and Mass Ratio of Unequal-Mass Protobinaries}
\label{massratio}

The vast majority of low (Sun-like) mass binaries are unequal-mass
systems (\ct{DuquennoyMayor1991}; \ct{Raghavan+2010}). The mass
ratio, defined as $q=M_2/M_1$ where $M_1$ and $M_2$ are the mass of 
the primary and secondary respectively, is one of the fundamental 
parameters that characterize the binaries. Its observed distribution 
provides important constraints on binary formation and evolution 
(see discussionn in \S~\ref{discussion}).

The mass ratio $q$ of protobinaries is expected to be affected by magnetic braking. This is because the change of $q$ depends on the specific angular momentum of the circumbinary material to be accreted relative to that of the binary. It is well known that high angular momentum material tends to accrete preferentially onto the less-massive secondary, which has a higher specific angular momentum than the primary, driving the mass ratio toward unity (\ct{BateBonnell1997}; \ct{Bate+2002}). As we have seen above, magnetic braking can remove the angular momentum of the circumbinary material efficiently. It is expected to weaken the tendency for preferential accretion onto the secondary. We show that it is indeed the case in Fig.~\ref{figq.25}. 
\begin{figure}
\epsscale{1.0}
\plotone{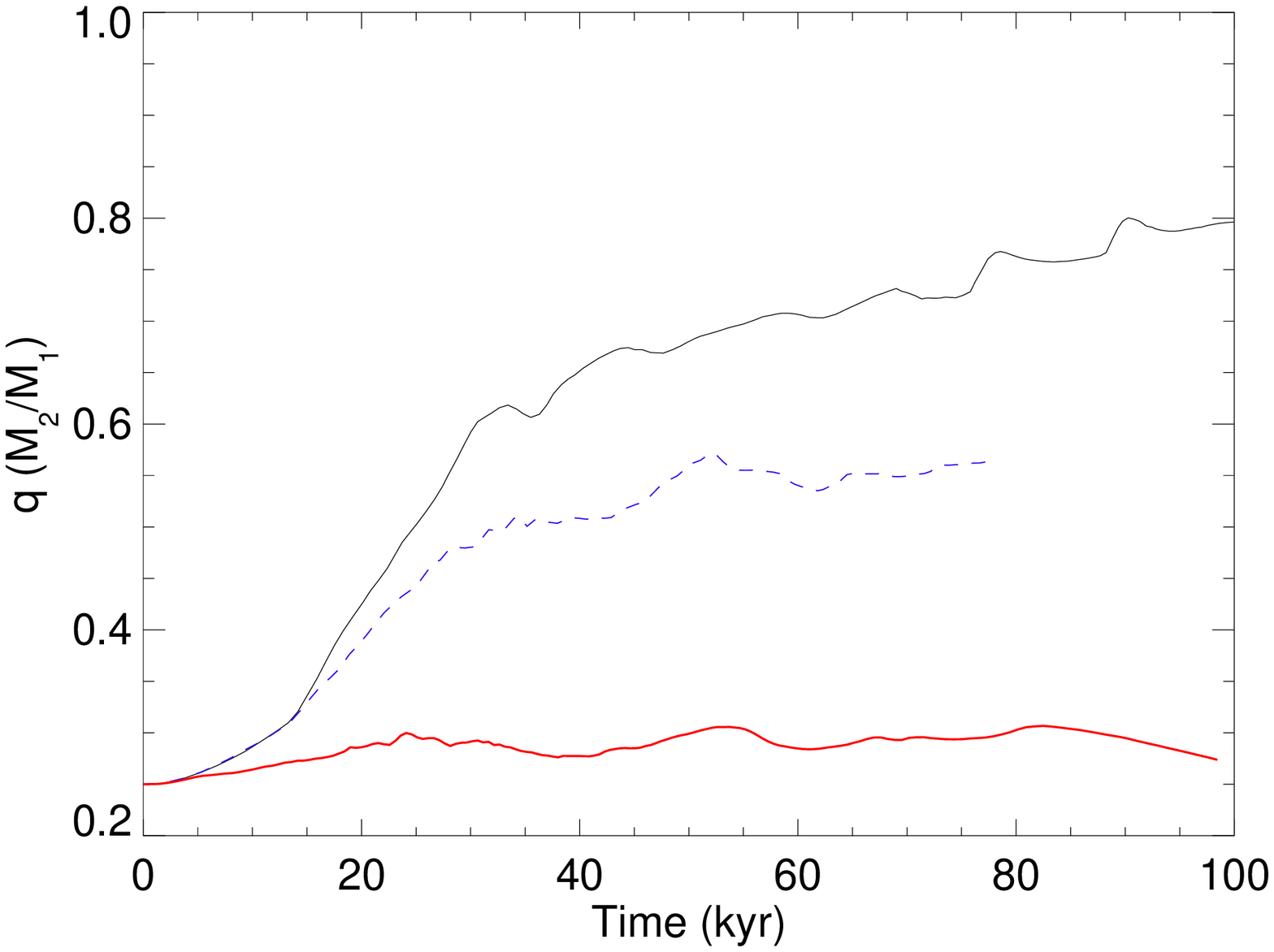}
\caption{Evolution of the protobinary mass ratio $q$ with time for envelopes that have different levels of magnetization, with $\lambda=\infty$ (HD, black solid), 16 (blue dashed), and 2 (red thick solid).}
\label{figq.25}
\end{figure}

Shown in Fig.~\ref{figq.25} are the mass ratios as a function of time
for three cases $\lambda=\infty$ (HD), 16 and 2. The initial
conditions are the same as those for the equal-mass cases discussed in
\S~\ref{result}, except that the mass and angular momentum inside the
radius $r_h$ of the power-law envelope are given to a binary system of
mass ratio $q=0.25$ rather than $q=1$. The mass ratio increases with
time in the HD case, in agreement with previous results. As expected,
the magnetic field slows down the increase in $q$ compared to the HD
case. Even a rather weak magnetic field of $\lambda=16$ has an
appreciable effect on the mass ratio evolution. For the more realistic
value of $\lambda=2$, the primary accretes much faster than the
secondary, by a factor of ~3-4, so that the mass ratio remains roughly
constant. The large contrast between the HD and $\lambda=2$ case
leaves little doubt that magnetic braking is an important factor to 
consider in understanding the mass ratio distribution of binaries. We 
will leave a comprehensive exploration of this issue to a future 
investigation. 

\section{Summary and Discussion}
\label{discussion}

%
% future work...different modes of binary formation, as well as eccentricity. 
%
%
We have carried out a set of idealized numerical experiments to demonstrate the effects of the magnetic field on protobinary evolution during the mass accretion phase. The protostellar envelope was idealized as a rotating, magnetized singular isothermal sphere. Its collapse onto a pre-existing pair of binary seeds was followed using an MHD version of the ENZO AMR hydro code that includes a sink particle treatment. We found that a magnetic field of the observed strength (corresponding to a dimensionless mass-to-flux ratio of a few) can remove, through magnetic braking, most of the angular momentum of the material that reaches the protobinary. The reduction in the angular momentum of the protobinary accretion flow has two important consequences: compared to the non-magnetic case, (1) the protobinary orbit becomes much tighter, and (2) the mass-ratio does not increase as fast with time for initially unequal mass systems. In addition, the magnetic field drastically changes the morphology and dynamics of the structures that surround the protobinary. It suppresses the formation of the familiar circumstellar and circumbinary disks in the non-magnetic case, as well as the spiral arms embedded in them. These structures are replaced by a bipolar magnetic-braking driven expanding regions, a dense, infalling, circumbinary pseudodisk in the equatorial region, and low-density, highly magnetized structures close to the protobinary that expand against the pseudodisk (the DEMS). We conclude that both the basic characteristics (such as mass ratio and separation) of the protobinaries and the environment in which binaries grow are strongly modified by a realistic magnetic field. 

%\subsection{Magnetic Braking-Driven Orbital Migration and Connection to Disk Formation}
  
The magnetic braking-driven inward migration of protobinaries may have
observable consequences. As mentioned in the introduction, the
distribution of binary orbital separation during the earliest, Class 0
phase of star formation may be different from those at later times. In
particular, there is tentative evidence for a ``desert'' free of Class
0 binaries with separation between $\sim 150-550$~AU (\ct{Maury+2010};
\ct{Enoch+2011}), which is not present in the Class I or later
phases. If confirmed by future observations, this gap must be filled
in by binaries from either outside the gap or interior to it. The
magnetic braking is an efficient way of shrinking the protobinary
orbit. It can in principle move some binaries born on wide orbits
outside the gap into the gap. Indeed, Offner et al. (2009, 2010)
  found that most of the binaries in their cluster formation
  simulations were born with relatively wide separations, as a result 
of radiative feedback, which tends to suppress close binary formation
through disk fragmentation; the orbits of such wide binaries could be 
tigthened by magnetic braking\footnote{Bate (2012) found that the
binaries in his radiation hydro simulations of cluster formation have
properties consistent with observations, indicating that magnetic
effects are not needed. It would be interesting to quantify how a magnetic 
field of the observed strength modifies the properties of the binaries
formed in a cluster environment.}.
A potential problem is that the braking may be so efficient during the
protobinary mass accretion phase (when a massive, slowly rotating
envelope is still present) that the binary separation would move
quickly through the gap and pile up below the resolution of the
current generation of millimeter/submillimeter interferometers ($\sim
50-100$~AU). If this is the case, one would expect to find an
over-abundance of relatively close protobinaries with separation
$\lesssim 100$~AU that may be uncovered by ALMA and JVLA. The same pileup at small separations would also occur if the binaries born on tight orbits interior to the gap are kept from expanding into the gap by magnetic braking. 

The problem of potentially turning most wide binaries into close
binaries is related to the magnetic braking catastrophe in disk formation (Galli et al. 2006). Both semi-analytic arguments and numerical simulations have shown that, in magnetized laminar dense cores of $\lambda$ of a few, the formation of a rotationally supported disk is  suppressed by magnetic braking in the ideal MHD limit (\ct{Allen+2003}; \ct{Galli+2006}; \ct{PriceBate2007}; \ct{MellonLi2008}; \ct{HennebelleFromang2008}; \ct{Seifried+2011}; \ct{DappBasu2011}). The magnetic braking
must be weakened somehow in order to form both large-scale disks and
wide binaries. In the context of disk formation, there are several
proposed mechanisms for weakening the magnetic braking, including the
misalignment between the magnetic and rotation axes (\ct{Joos+2012}),
turbulence (\ct{Santos-Lima+2012}; \ct{Seifried+2012}; \ct{Myers+2012}),
and the depletion of the slowly rotating protostellar envelope that
acts to brake the disk, either by outflow stripping (\ct{MellonLi2008}) or accretion (\ct{Machida+2010}). The last possibility is
particularly intriguing, because it implies a rapid growth of the
rotationally supported disk during the transition from the deeply
embedded Class 0 phase to the more revealed Class I phase that can be
observationally tested. Similarly, the depletion of the protobinary
envelope and the associated weakening of magnetic braking may enable
the separation of the protobinaries to grow quickly during the Class
0-Class I transition. If this is the case, the orbits of wide binaries
may first shrink during the Class 0 phase due to efficient magnetic
braking and then re-expand as the protobinary envelope depletes. This
and other possibilities for protobinary migration should be testable
with high-resolution millimeter/submillimeter interferometric
observations, especially using ALMA and JVLA. 
%
% suppression of disks...disk fragmentation and binary migration by circumbinary interaction. 
%
% effects on Bate (2012) simulations...any gain?

Our calculations demonstrated that magnetic braking is important for
the evolution of not only the binary separation, but also the mass
ratio $q$. \citet{Raghavan+2010} found a roughly flat distribution
for $q$ between 0.2-0.95, with a valley below 0.2, and three spikes
around 0.25, 0.35 and 1.0 (see the left panel of their Fig.~16). It is
tempting to attribute the last spike at $q\sim 1$ to the preferential
accretion of high specific angular momentum material onto the
lower-mass component found in hydrodynamical simulations (\ct{BateBonnell1997}; \ct{Bate+2002}). However, the spike around $q\sim 1$
is just one of the three spikes in \citeauthor{Raghavan+2010}'s data, 
and it does not show up in the sample of \citet{DuquennoyMayor1991} 
at all (see their Fig.~10). In any case, there are many more systems 
with mass ratio $q$ well below unity than close to unity. A mechanism 
must be found to prevent the majority of the low-mass ratio
protobinaries from becoming equal-mass systems. 
%{\bf In a cluster context, the turublent-radiation simulation by
%\ct{Bate2012} reproduces a flat mass ratio distribution for stars
%with mass $\gtrsim 0.5\msun$.} 
The magnetic braking highlighted in this paper is one such mechanism. It is a key factor to consider in understanding binary formation and evolution in dense cores that are observed to be significantly magnetized. 

\acknowledgments
We thank P. Arras, S. Offner, K. Kratter, C. Matzner, and A. Maury for useful discussion and P. Wang for advice on the ENZO code. This work is supported in part by NASA NNX10AH30G.

\end{document}